\documentclass[11pt]{article}
\usepackage{moriond,epsfig}
\usepackage{amsmath}

\def\be{\begin{equation}}
\def\ee{\end{equation}}
\def\bea{\begin{eqnarray}}
\def\eea{\end{eqnarray}}

\begin{document}
\begin{flushright}
   Cavendish-HEP-05-10
\end{flushright}
\vspace*{4cm}
\title{A Closer Look at the Analysis of NLL BFKL}

\author{Jeppe R. Andersen}

\address{Cavendish Laboratory, University of Cambridge, Madingley Road, CB3
  0HE, Cambridge, UK} \maketitle\abstracts{The initial analyses of the
  next-to-leading logarithmic corrections to the BFKL kernel were very
  discouraging. Encouraged by the success of new methods in the analysis of
  the BFKL equation at full NLL accuracy we demonstrate in this talk how some
  of the initial conclusions were based on a breakdown of the tools used in
  the analysis rather than the framework itself.  }
\section*{Introduction}
\label{sec:introduction}

The Balitsky--Fadin--Kuraev--Lipatov{}\cite{FKL} (BFKL) framework
systematically resums the class of logarithms originating from the kinematics
that dominate the total cross section in the Regge limit of scattering
amplitudes, where the centre of mass energy $\sqrt{\hat{s}}$ is large and the
momentum transfer $\sqrt{-\hat{t}}$ is fixed. In this limit the scattering of
two gluons $p_A p_B\to p_{A'}p_{B'}$ will be dominated by multi--particle
production leading to final states described by momenta
$k_0=p_{A'},k_1,\ldots, k_n,k_{n+1}=p_{B'}$ satisfying
\begin{align}
\label{eq:multiregge}
  s\gg2 k_{i-1}k_i\gg t_i=q_{i}^2, q_i=p_A-\sum_{r_0}^{i-1}k_r,
  \prod_{i=1}^{n+1}s_i=s\prod_{i=1}^n \mathbf{k}_i^2,
  k_{\perp}^2=-\mathbf{k}^2, |k_{i\perp}|\simeq|p_{A'\perp}|
\end{align}
The Regge limit is therefore suitable for describing the production of
multiple hard partons from e.g.~gluon scattering (and in fact the
large-rapidity limit of any process that includes a $t$-channel gluon
exchange). We will in this talk focus entirely on processes within the
multi-Regge kinematics of Eq.~(\ref{eq:multiregge}). In this limit the
partonic cross section can be approximated by
\begin{eqnarray}
\label{cross--section1}
\hat\sigma(\Delta) &=&\int 
\frac{d^2 {\bf k}_a}{2 \pi{\bf k}_a^2}
\int \frac{d^2 {\bf k}_b}{2 \pi {\bf k}_b^2} ~\Phi_A({\bf k}_a) 
~f \left({\bf k}_a,{\bf k}_b, \Delta\right)~\Phi_B({\bf k}_b),
\end{eqnarray}
where $\Phi_{A,B}$ are the impact factors characteristic of the particular
scattering process, and $f\left({\bf k}_a,{\bf k}_b,\Delta\right)$ is the
gluon Green's function describing the interaction between two Reggeised
gluons exchanged in the $t$--channel with transverse momenta ${\bf k}_{a,b}$,
spanning a rapidity interval of length $\Delta$. The leading and
next-to-leading logarithmic contributions to this gluon Green's function can be
resummed by solving the BFKL equation to the required accuracy
\begin{eqnarray}
\label{eq:BFKLeqn}
\omega \ f_\omega\! \left({\bf k}_a,{\bf k}_b\right) &=& \delta^{(2+2\epsilon)} 
\left({\bf k}_a-{\bf k}_b\right) + \int d^{2+2\epsilon}{\bf k} ~
\mathcal{K}\!\left({\bf k}_a,{\bf k}+{\bf k}_a\right) \ f_\omega\!\left({\bf k}+{\bf k}_a,{\bf k}_b \right),
\end{eqnarray}
where $w$ is the Mellin-conjugated variable to $\Delta$, and the BFKL kernel
$\mathcal{K}\!\left({\bf k}_a,{\bf k}+{\bf k}_a\right)$ is presently known to
next-to-leading logarithmic accuracy.

\section{Solutions of the BFKL equation}
\label{sec:solut-bfkl-equat}
The solution to integral equations of the form in Eq.~\eqref{eq:BFKLeqn} can
be written in terms of the eigenfunctions $\phi_i(k_a)$ and eigenvalues
$\lambda_i$ as
\begin{align}
  \label{eq:gensoln}
  f_{\omega}(k_a,k_b)=\sum_i\!\!\!\!\!\!\!\!\int\frac{\phi_i(k_a)\ \phi_i^*(k_b)}{\omega-\lambda_i}
\end{align}
leading to
\begin{align}
  \label{eq:gensolndelta}
  f(k_a,k_b,\Delta)=\sum_i\!\!\!\!\!\!\!\!\int \frac 1 {2\pi i}\ e^{\Delta
    \lambda_i}\ \phi_i(k_a)\ \phi_i^*(k_b)
\end{align}

\subsection{Leading Logarithmic Accuracy}
\label{sec:lead-logar-accur}

At leading logarithmic accuracy the BFKL kernel is conformal invariant, since
the running of the coupling only enters at higher logarithmic orders. The
eigenfunctions of the angular averaged kernel are of the form
${k^2}^{(\gamma-1)}, \gamma=1/2+i \nu$, which means that to this accuracy,
the BFKL evolution can be solved analytically, with the transverse momentum
of emitted gluons integrated to infinity, by analysing the Mellin transform
of the kernel. One finds
\begin{eqnarray}
  \label{eq:LLev1}
  \int\mathrm{d}^{D-2}\mathbf{k}\ \mathcal{K}^{\mathrm{LL}}\!\left({\bf
    k}_a,{\bf k}\right)\ \left({\mathbf{k}^2}
  \right)^{\gamma-1}=\frac{\alpha_s N}\pi \chi^{\mathrm{LL}}(\gamma)
  \left(\mathbf{k}_a^2\right) ^{\gamma-1},
\end{eqnarray}
with $N$ being the number of colours and 
\begin{eqnarray}
  \label{eq:LLev2}
  \chi^{\mathrm{LL}}(\gamma)=2\psi(1)-\psi(\gamma)-\psi(1-\gamma),\qquad \psi(\gamma)=\Gamma'(\gamma)/\Gamma(\gamma).
\end{eqnarray}
Since both the eigenfunctions and eigenvalues are known, the angular averaged
gluon Green's function can now be obtained according to
Eq.~\eqref{eq:gensolndelta} as 
\begin{eqnarray}
  \label{eq:angular_avg_f}
  \bar f(k_a,k_b,\Delta)=\frac 1 {k_b^2}\ \int_{\frac 1 2 - i \infty}^{\frac 1 2 + i
  \infty} \frac{\mathrm{d}\gamma}{2\pi i}\ e^{\Delta
  \omega^{\mathrm{LL}}(\gamma)}\left(\frac{k_b^2} {k_a^2} \right)^{\gamma},
\end{eqnarray}
where
\begin{eqnarray}
  \label{eq:omega_LL}
\omega^{\mathrm{LL}}(\gamma)\equiv  \int\mathrm{d}^{D-2}\mathbf{k}\ \mathcal{K}^{\mathrm{LL}}\!\left({\bf
    k}_a,{\bf k}\right)\ \left(\frac{\mathbf{k}^2}{\mathbf{k}_a^2}\right)^{\gamma-1}
=\frac{\alpha_s(k_a^2)N}\pi \chi^{\mathrm{LL}}(\gamma).
\end{eqnarray}
We stress that at LL the coupling is formally fixed, and so the
regularisation scale is completely arbitrary. 
\begin{figure}[tbp]
  \centering
  \epsfig{width=10cm,file=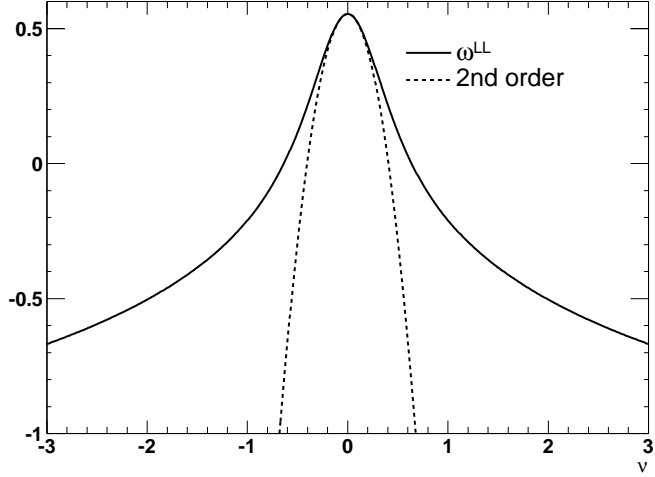}
  \caption{$\omega^{\mathrm{LL}}(\frac 1 2+i \nu)$ and the second order
  Taylor polynomial around $\nu=0$.}
  \label{fig:omegaLL}
\end{figure}
In Fig.~\ref{fig:omegaLL} we have plotted $\omega^{\mathrm{LL}}(\frac 1 2+i
\nu)$ for $\alpha_s=0.2$ and it is seen that there is a maximum at $\nu=0$.
Therefore, the behaviour of the gluon Green's function in the limit of
$\Delta\to\infty$ is determined by the value of $\omega^{\mathrm{LL}}(\frac 1
2)=4\ln 2\alpha_s N/\pi$. A saddle point approximation based on the second
order Taylor polynomial around $\nu=0$ will correctly describe the asymptotic
exponential growth in $\Delta$, since the polynomial attains the correct
value at the maximum.

\subsection{Next-to-Leading Logarithmic Accuracy}
\label{sec:next-lead-logar}
\begin{figure}[tbp]
  \centering
  \epsfig{width=10cm,file=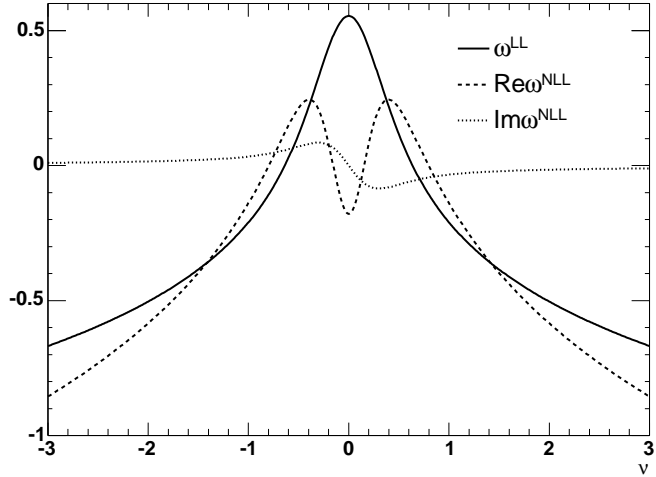}
  \caption{$\omega^{\mathrm{LL}}(\frac 1 2 + i \nu)$ and
  $\omega^{\mathrm{NLL}}(\frac 1 2 + i \nu)$ for $\alpha_s=0.2$, and
  $N=n_f=3$.} 
  \label{fig:omegaNLL}
\end{figure}

When trying to extend this analysis to NLL accuracy one is immediately faced
with the complications introduced by the breaking of the conformal invariance
by the running coupling terms. This effect will necessarily change the
eigenfunctions, and thus far the eigenfunctions for the full NLL kernel in
QCD have not been constructed. Traditionally, the kernel at NLL has been
studied using the projection on the Born level eigenfunctions as in
Eq.~\eqref{eq:omega_LL}. One finds~\cite{Fadin:1998py}
\begin{align}
  \label{eq:NLL_kernel_on_LL_ef}
  \begin{split}
    \omega^{\mathrm{NLL}}(\gamma)\equiv &\int\mathrm{d}^{D-2}\mathbf{k}\ 
    \mathcal{K}^{\mathrm{NLL}}\!\left({\bf k}_a,{\bf
        k}\right)\left(\frac{\mathbf{k}^2} {\mathbf{k}_a^2}
    \right)^{\gamma-1}\\=&\frac{\alpha_s(\mathbf{k}_a^2)N}\pi
    \left(\chi^{\mathrm{LL}}(\gamma)+\chi^{\mathrm{NLL}}(\gamma)\frac{\alpha_s(\mathbf{k}_a^2)N}{\pi}
    \right)
\end{split}
\end{align}
with
\begin{align}
  \label{eq:chi_NLL}
  \begin{split}
    \chi^{\mathrm{NLL}}(\gamma)=&-\frac 1 4\Big[\left(\frac {11}3-\frac 2
        3\frac {n_f}N\right)\frac 1
        2\left(\chi^{\mathrm{LL}}(\gamma)-\psi'(\gamma)+\psi'(1-\gamma) 
\right)\\
&-6 \zeta(3) +\frac{\pi^2\cos(\pi\gamma)}{\sin^2(\pi\gamma)(1-2\gamma)}
\left(3+\left(1+\frac{n_f}{N^3}\right)
\frac{2+3\gamma(1-\gamma)}{(3-2\gamma)(1+2\gamma)}
\right)\\
&-\left(\frac{67}9-\frac{\pi^2}3-\frac{10}9\frac{n_f}N
\right)\chi^{\mathrm{LL}}-\psi''(\gamma)-\psi''(1-\gamma) -
        \frac{\pi^3}{\sin(\pi\gamma)} + 4 \phi(\gamma)
\Big],
\end{split}
\end{align}
where
\begin{align}
  \label{eq:phi}
  \begin{split}
    \phi(\gamma)=&-\int_0^1\frac{\mathrm{d}x}{1+x}(x^{\gamma-1}+x^{-\gamma})
    \int_x^1 \frac{\mathrm{d}t}t\ln(1-t)\\
    =&\sum_{n=0}^\infty(-1)^n\left[\frac{\psi(n+1+\gamma)-\psi(1)}{(n+\gamma)^2}
    + \frac{\psi(n+2-\gamma)-\psi(1)}{(n+1-\gamma)^2}
\right].
\end{split}
\end{align}
An approximation to the gluon Green's function at NLL can then be constructed
by use of $\omega^{\mathrm{NLL}}$ in place of $\omega^{\mathrm{LL}}$ in
Eq.~(\ref{eq:angular_avg_f}). 
We have in Fig.~\ref{fig:omegaNLL} plotted the real and imaginary part of
$\omega^{\mathrm{NLL}}$ compared with $\omega^{\mathrm{LL}}$ for
$\alpha_s=0.2$. The double hump structure of the real part of
$\omega^{\mathrm{NLL}}$ is potentially a disaster for the gluon Green's
function. At asymptotically large $\Delta$ the behaviour of the gluon Green's
function is determined by the position and value of the maxima of
$\omega^{\mathrm{NLL}}$ only. Since there are two such distinct maxima located at
$\gamma_1=1/2-i\nu_0, \gamma_2=1/2+i\nu_0$, the asymptotic estimate of the
NLL gluon Green's function based on the LL eigenfunctions will have an
oscillatory behaviour in $\ln k_a/k_b$. This is a problem of matching to the
DGLAP evolution, and is a problem strictly outside the Regge kinematics.

The initial observation of the large NLL corrections was based on the
difference $\omega^{\mathrm{LL}}-\omega^{\mathrm{NLL}}$ evaluated at $\nu=0$.
Indeed, for reasonable values of the coupling, the real part of
$\omega^{\mathrm{NLL}}$ is even negative at this point. However, this is not
what determines the intercept.  If indeed the solution of the BFKL equation
at NLL was obtained using $\omega^{\mathrm{NLL}}$ of
Eq.~\eqref{eq:NLL_kernel_on_LL_ef}, then the asymptotic intercept at NLL
would again be determined by the maximum value attained by the real part of
$\omega^{\mathrm{NLL}}(\gamma)$ along the contour $\gamma=1/2+i \nu$. We see
from Fig.~\ref{fig:omegaNLL} that this maximum value is roughly halved
compared to the LL asymptotic intercept.
%% Asymptotic intercept determined by the maximum of the real part.

However, even within the Regge kinematics, this analysis leads to severe
problems in the very limit, where the resummed logarithmic terms are meant to
dominate the scattering matrix. The non-zero imaginary part of
$\omega^{\mathrm{NLL}}(\frac 1 2 + i \nu_0)$ leads to oscillations with
increasing rapidity for any choice of $k_a$ and $k_b$! This clearly signals a
breakdown of the approach in the very limit it is meant to describe.

The solution to this apparent problem is the realisation that what one has
studied with this method is indeed not the true solution to the BFKL equation
at NLL. The LL eigenfunctions are \emph{not} the eigenfunctions at NLL (for
non-zero $\beta_0$). Indeed, the only contribution to the troublesome
imaginary part of $\omega^{\mathrm{NLL}}(\gamma)$ for $\gamma=1/2+i \nu$
stems from the term $-\psi'(\gamma)+\psi(1-\gamma)$ in
Eq.~\eqref{eq:chi_NLL}, which contributes to $\omega^{\mathrm{NLL}}$ with a
factor proportional to $\beta_0$ (that is, it vanishes in the limit where the
LL eigenfunctions diagonalises the NLL kernel). It was observed already in
Ref.\cite{Fadin:1998py} that this part of the NLL corrections is the only one
that is not symmetric under $\gamma\leftrightarrow 1-\gamma$, and that if one
expands the kernel on the LL eigenfunctions rescaled by a square root of the
coupling,
i.e.~$(k^2)^{\gamma-1}\left(\frac{\alpha_s(k^s)}{\alpha_s(\mu^2)}\right)^{-1/2}$
then these and only these terms would disappear from Eq.~\eqref{eq:chi_NLL}.
What was perhaps not realised is that since this is the only contribution to
the imaginary part of $\omega^\mathrm{NLL}$, this would simultaneously cure
the problem of oscillations within the Regge-limit.  It should be emphasised
though that these rescaled functions still are not the true eigenfunctions at
NLL, but it is straightforward to check numerically that the ``off-diagonal
elements'' of $\omega^{\mathrm{NLL}}$ (i.e. those obtained with a different
$\gamma$ for $k_a$ and $k$ in Eq.~\eqref{eq:NLL_kernel_on_LL_ef}) are far
smaller in this case than when using the pure LL eigenfunctions. With the
advance of new approaches to the solution of the BFKL equation at full NLL
accuracy\cite{Andersen:2003an,Andersen:2003wy,Andersen:2004uj,Ciafaloni:2003rd}
it has also been possible to check explicitly how well the two approximations
compare with the full solution. We find that the approximation using the
rescaled LL eigenfunctions is much closer to the true solution than the one
using the pure LL ones. It should be noted that a saddle point approximation
based around $\nu=0$ would have to use an extremely large order approximation
in order to describe correctly the asymptotic intercept. Even the 16th order
Taylor polynomial would fail to reach the maximum value for
$\omega^{\mathrm{NLL}}$ in Fig.~\ref{fig:omegaNLL}. It would be far better to
base a saddle point approximation around $\nu_0$.

\begin{figure}[htbp]
  \centering
  \epsfig{width=10cm,file=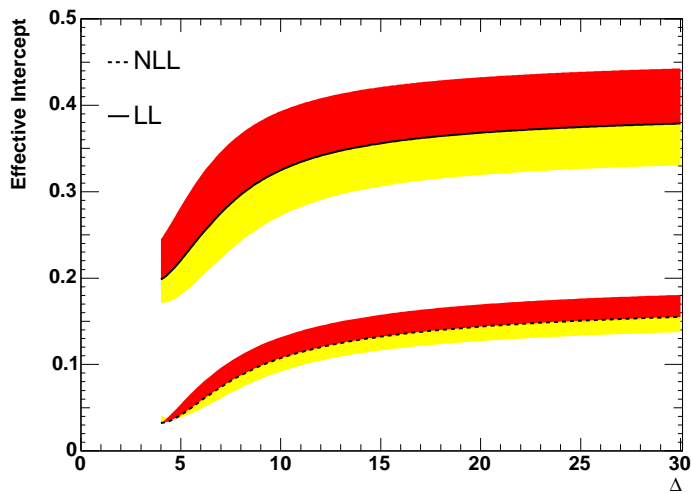}
  \caption{The effective intercept (of
    $f(k_a=20\mathrm{GeV},k_b=30\mathrm{GeV},\Delta)$ with
    $\alpha_s(30\mathrm{GeV})=0.15$), as a function of the rapidity at LL
    (upper line) and NLL (lower line).  The uncertainty due to a
    renormalisation scale variation of a factor of two is indicated by the
    colour band.}
  \label{fig:intercept}
\end{figure}

Using the guess obtained from these rescaled eigenfunctions it is possible to
calculate the intercept as the logarithmic derivative of the gluon Green's
function for fixed $k_a$ and $k_b$, as a function of the rapidity. This is depicted on
Fig.~\ref{fig:intercept}. We see that although the NLL correction amounts to
roughly a factor of two, it is stable. Also, it should be remembered that the
study of both the LL and NLL intercept here has been performed without
constraining the phase space of the BFKL resummation to such which is
attainable at a given collider. The effects of such a constrain are known to
be
large\cite{Andersen:2003gs,Orr:1997im,Orr:1998hc,Orr:1998ps,Andersen:2001ja,Andersen:2001kt}
and reduce the LL evolution significantly (and presumably the NLL intercept to a
slightly lesser extent).

\section*{Conclusions}
We have shown that the NLL corrections to the BFKL intercept are large but
stable within the Regge kinematics. The instability with respect to the
evolution in rapidity observed in initial analyses is a direct result of
using the conformal set of leading log eigenfunctions as if they were
eigenfunctions at NLL. Indeed, for $\beta_0=0$ this instability disappear
even in this analysis, and this the case also for non-zero $\beta_0$ if one
applies rescaled eigenfunctions. Although these still do not diagonalise the
kernel at NLL, the results obtained using the rescaled eigenfunctions
describe the full solution obtained numerically much better than the
approximation obtain using just the LL eigenfunctions. In the conformal limit
of $\beta_0=0$, where the NLL corrections can be studied exactly using the
projections on the LL eigenfunctions, any modification of the NLL kernel to
match better to the DGLAP region (like the ones of
Ref.\cite{Salam:1998tj,Vera:2005jt}) must move the position of the maximum of
$\omega^{\mathrm{NLL}}$ to $\nu=0$ while not changing the maximum value
itself significantly (since this would lead to a change in the asymptotic
intercept obtained within the Regge kinematic).

\section*{Acknowledgments}
I would like to thank Agust\'in Sabio Vera for lively discussions. This
research was supported by PPARC (postdoctoral fellowship PPA/P/S/2003/00281).

\end{document}